\documentclass[superscriptaddress, preprint, apl]{revtex4}

\usepackage{graphicx}   

\newcommand{\figref}[1]{\figurename~\ref{#1}}

\newcommand{\meter}[1][]{\ifx|#1|\unit{m}\else\unit[#1]{m}\fi}
\newcommand{\hertz}[1][]{\ifx|#1|\unit{Hz}\else\unit[#1]{Hz}\fi}
\newcommand{\fm}[1][]{\ifx|#1|\unit{fm}\else\unit[#1]{fm}\fi}
\newcommand{\fluence}[1][]{\ifx|#1|$\unit{mJ/cm^2}$\else$\unit[#1]{mJ/cm^2}$\fi}

\begin{document}

\title{Coherent phonon dynamics at the martensitic phase transition of Ni$_2$MnGa}

\author{S. O. Mariager}
\email{simon.mariager@psi.ch}
\address{Swiss Light Source, Paul Scherrer Institut, 5232 Villigen, Switzerland}
\author{A. Caviezel}
\address{Swiss Light Source, Paul Scherrer Institut, 5232 Villigen, Switzerland}
\author{P. Beaud}
\address{Swiss Light Source, Paul Scherrer Institut, 5232 Villigen, Switzerland}
\author{C. Quitmann}
\address{Swiss Light Source, Paul Scherrer Institut, 5232 Villigen, Switzerland}
\author{G. Ingold}
\address{Swiss Light Source, Paul Scherrer Institut, 5232 Villigen, Switzerland}

\date{27-04-2012}

\begin{abstract} We use time-resolved optical reflectivity to study the laser stimulated dynamics in the
magnetic shape memory alloy Ni$_2$MnGa. We observe two coherent
optical phonons, at 1.2~THz in the martensite phase and at 0.7~THz
in the pre-martensite phase, which we interpret as a zone-folded
acoustic phonon and a heavily damped amplitudon respectively. In the
martensite phase the martensitic phase transition can be induced by
a fs laser pulse on a timescale of a few ps.
\end{abstract}

\maketitle In magnetic shape memory Heusler alloys the combination
of ferromagnetism and a structural martensitic phase transition
leads to large magnetic field induced strains, reaching up to 10\%
in the prototype Ni$_2$MnGa \cite{Sozinov2002}. While practical
applications already exist, both the structure of the martensite
phase (MT) \cite{Kaufmann2010} and the microscopic origin of the
phase transition remain under discussion \cite{Opeil2008,
Uijttewaal2009}. Upon cooling Ni$_2$MnGa undergoes two structural
phase transitions. At T$_{PMT} \approx$ 260~K a pre-martensitic
transition leads to a gradual modulation of the high temperature
cubic austenite phase (AUS) of L2$_1$ symmetry to the pre-martensite
phase (PMT). The subsequent first order martensitic transition
occurs at T$_{MT} \approx$ 220~K. Through both structural
transitions Ni$_2$MnGa remains ferromagnetic (T$_C \approx$ 380~K).

The PMT occurs due to a softening of the [$\zeta\zeta$0] acoustic
phonon mode at  $\zeta_0$ = 1/3. The resulting structure arises from
a periodic modulation of the positions of the atoms within the cubic
unit cell tripling the unit cell along \{110\}
\cite{Zayak2003,Brown2002,Uijttewaal2009}. The structure of the MT
is however debated, in part because different stoichiometries lead
to different structures \cite{Lanska2004}. The reported structures
can roughly be summarized as tetragonal or orthorhombic distortions
of the AUS unit cell, combined with a 5 or 7 fold distortion along
\{110\}. The latter distortion has been described both as a periodic
modulation and a shuffling of atomic layers \cite{Pons2005}. A
possible explanation was given recently when the MT structure of
Ni$_2$MnGa was described as an adaptive phase
\cite{Kaufmann2010,Khachaturyan1991}, where nanotwinning occurs
along \{110\} in order to minimize the elastic energy at the
cubic/tetragonal interface. As a result there are 12 tetragonal
twins \cite{Jakob2007} and the 5 and 7 fold distortions are
$(3\overline{2})_2$ and $(5\overline{2})_2$ in Zhadonov notation
\cite{Ustinov2009}. Also the driving forces of the martensitic
transition are not fully understood on the microscopic level. The
electronic stabilization appears to be a pseudogap 0.3~eV below the
Fermi level. The pseudogap has both been measured with
photo-emission \cite{Opeil2008} and calculated theoretically
\cite{Ayuela2002, Entel2010}. It forms by redistribution of minority
spin Ni-3d states and can be understood as a band Jahn-Teller
effect. In addition it has recently been stressed that the combined
magnetic and vibrational excitations must be included in order to
reproduce the full phase diagram of Ni$_2$MnGa
\cite{Uijttewaal2009}, while coupling between acoustic and optical
phonon modes in the AUS can account for the wavevectors related to
the modulation \cite{Zayak2005}.

Motivated by the fact that fs laser pulses can excite coherent
optical phonons \cite{Ishioka2010} and induce structural phase
transitions through the release of Jahn-Teller distortions
\cite{Beaud2009} we studied the phonon modes and structural dynamics
in Ni$_2$MnGa by all optical time-resolved reflectivity. In this
letter we show that low fluence excitations of the different phases
excite two distinct coherent optical phonons in the PMT and MT
respectively. Secondly the martensitic phase transition can be
induced, leading to signatures of the PMT appearing after a few ps.

The samples were a stoichiometric Ni$_2$MnGa crystal
($10\times10\times1$~mm$^3$) and non-stochiometric
Ni$_{50}$Mn$_{28.3}$Ga$_{21.7}$ crystal ($4\times4\times1$~mm$^3$)
(both from AdaptaMat Ltd.). Both crystals were cut and polished with
\{001\} facets. The photoinduced change in reflectivity was
investigated in an optical pump-probe setup with a time resolution
of $\sim 80$~fs. The 50~fs 800~nm pump and probe pulse were
generated with a 2~kHz Ti-Saphire laser system. The pump and probe
pulses were cross polarized and incident along the [001] surface
normal with the probe polarization aligned to the [100] crystal
axis. The pump was focused to $500\times500~\mu m^2$ and the probe
was a factor of three smaller to ensure homogenous excitation of the
probe region. The sample was mounted in a cryostat.

\begin{figure} [b]
\includegraphics[scale=1]{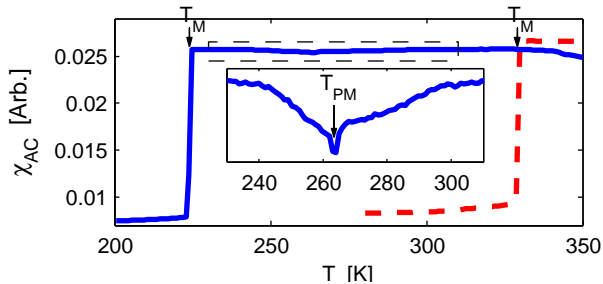}
\caption{(Color Online) AC magnetization curves for the
stoichiometric (blue, solid) and non-stoichiometric (red, dashed)
Ni$_2$MnGa crystals. The inset highlights the pre-martensite region
and correspond to the dashed box in the main figure.}
\label{figure1}
\end{figure}

To verify the existence of the MT the magnetic AC susceptibility was
measured as a function of sample temperature. The measurements were
performed with a Quantum Design PPMS with a driving field of 10~Oe
and no applied static field. The results are independent of
frequency between 13 and 3500~Hz and shown in \figref{figure1} for
34~Hz. The Martensitic transition is visible as a sharp change in
susceptibility at T$_{MT}$ = 224~K and T$_{MT}$ = 329~K for the two
crystals respectively. The change in susceptibility is due to the
lower crystal symmetry and higher magneto-crystalline anisotropy of
the MT. For the stoichiometric crystal the non-magnetic
pre-martensitic transition is visible in the inset at T$_{PMT}$ =
264~K. The magnetic susceptibilities in \figref{figure1} are
consistent with the literature \cite{Sozinov2002,Opeil2008}.

\begin{figure} [t]
\includegraphics[scale=1]{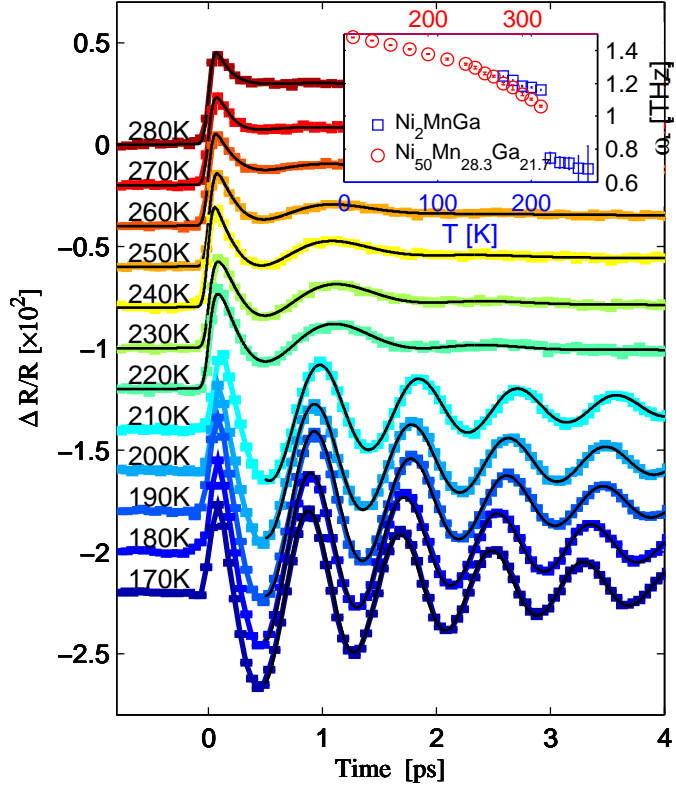}
\caption{(Color online) Time resolved reflectivity on Ni$_2$MnGa
obtained with a laser pump power of 1~mJ/cm$^2$. Black lines are
fits to the data. The insert shows the frequencies in both phases as
a function of temperature (blue squares), and also includes data
from the non-stoichiometric crystal (red circles). } \label{figure2}
\end{figure}

In \figref{figure2} we show the relative change in reflectivity as a
function of time after laser excitation for the stoichiometric
crystal at temperatures between 170 and 280~K. In the MT (T $<$
215~K) a strong oscillation with a low damping is seen in the
transient reflectivity. This is the signature of a coherent optical
phonon \cite{Ishioka2010}. Above 210~K the signal is markedly
different. At intermediate temperatures (220 - 260~K) an oscillation
is still visible, but the frequency is lower and the stronger
damping quenches the oscillation after a single period. As the
temperature is increased the damping gets stronger and at 270~K the
oscillation is gone. Here the change in reflectivity is reduced to
the incoherent response of a metal, due to the heating and
thermalization of the electron gas and the lattice
\cite{Hohlfeld2000}. The absence of coherent phonons in the AUS is
not surprising. In opaque materials fs laser pulses mainly excite
symmetry conserving A$_{1g}$ modes by displacive excitation, and no
such modes exist in the AUS of Ni$_2$MnGa \cite{Zayak2005}. From
\figref{figure2} it is obvious that transient reflectivity probes
the phase diagram of Ni$_2$MnGa. The transition temperatures T$_M$
and T$_{PM}$ agree with those in \figref{figure1} and the slight
shift to lower temperatures is due to the hysteresis of the first
order phase transition. The susceptibility was measured during
heating and the reflectivity during cooling. We find the same
phonons in the non-stoichiometric crystal, though the damping in the
PMT is stronger. This is due either to the higher temperature or the
stoichiometry.

To extract the phonon frequencies we fit the data from the MT with a
single damped cosine function $\Delta R_{osc} = A\cos(2\pi\omega
t+\Phi)e^{-t/\tau}$. To minimize the number of free parameters we
ignore the initial peak, which as explained below originates from
the heated electrons, by restricting the fit to t $>$ 0.5~ps. For
the PMT we fit the entire curve and account for the incoherent part
by using $\Delta R = \Delta R_{osc}+\Delta R_{e}$ with
\begin{equation}
\Delta R_{e}(t) =
a(1-e^{-t/\tau_{ee}})e^{-t/\tau_{ep}}+b(1-e^{-t/\tau_{ep}})e^{-t/\tau_{th}}
\label{equation1}
\end{equation}
Where $\tau_{ee}$ and $\tau_{ep}$ ($\approx 0.8~ps$) are the
electron-electron and electron-phonon relaxation times and
$\tau_{th}$ describes the subsequent cooling of the lattice. The
constants a and b determine the ratio between the electronic peak
and the thermal contribution. The fit function was convoluted with a
Gaussian accounting for the 80~fs time-resolution and the resulting
fits are shown in \figref{figure2}. In the inset the frequencies
$\omega$ are plotted as a function of temperature with error bars
given by 95\% confidence intervals from the fits. Because of the
high damping in the PMT phase of the non-stoichiometric crystal,
this frequency could not be extracted at a fluence of 1~mJ/cm$^2$.
In both phases the frequency softens with increasing temperature.
Thus thermal expansion rather than anharmonic effects dominates the
temperature dependence of the frequencies. We note that due to the
long wavelength of the optical light compared to the crystals
lattice constant the momentum transfer is $\sim0$ and the phonon
frequencies are measured at the $\Gamma$ point of the Brillouin
zone.

The acoustic phonon dispersion curves of Ni$_2$MnGa have been
measured by inelastic neutron scattering \cite{Zheludev1996}, but we
are not aware of any experimental results on optical phonons. The
phonon-dispersion curves have been calculated for both the AUS, PMT
and MT \cite{Uijttewaal2009,Zayak2003}. For the non-modulated
lattices the lowest optical phonon modes have frequencies above
4~THz in the zone-center and cannot explain the observed
frequencies. The structural modulations in the PMT and MT however
multiply the unit cell along \{110\}, leading to a reduced Brillouin
zone and zone-folded phonons. Since the measured phonon frequencies
in both phases are lower than the calculated frequencies of the
non-modulated unit cells, the phonon modes must be related to the
modulation. For the PMT low lying phonon modes consistent with our
measurements have indeed been predicted by considering the complete
tripled unit cell \cite{Uijttewaal2009}.

The crystal structure of the PMT is a periodic modulation of the
cubic unit cell along \{110\} with a period of 3a$/\sqrt{2}$. The
specific phonon mode corresponding to the oscillation of the
modulation amplitude is termed an amplitudon. A factor group
analysis \cite{Kroumova2003} using the modulated structure of the
PMT \cite{Brown2002} confirms the existence of a Raman active
symmetry conserving A$_{1g}$ mode corresponding to the amplitudon.
The phase of the phonon is difficult to determine due to the strong
damping and overlap in time with the electronic peak. We find that
$\phi = 0.59\pm.15$~rad at 1mJ/cm$^2$ and goes to zero at higher
pump fluences. This indicates a displacive excitation similar to the
well known laser excitation of the A$_1g$ mode of Bi
\cite{Sokolowski-Tinten2003}. Based on the low frequency, the factor
group analysis and the displacive character of the excitation, we
suggest that the optical phonon mode observed in the PMT is the
amplitudon of the modulated unit cell.

\begin{figure} [t]
\includegraphics[scale=1]{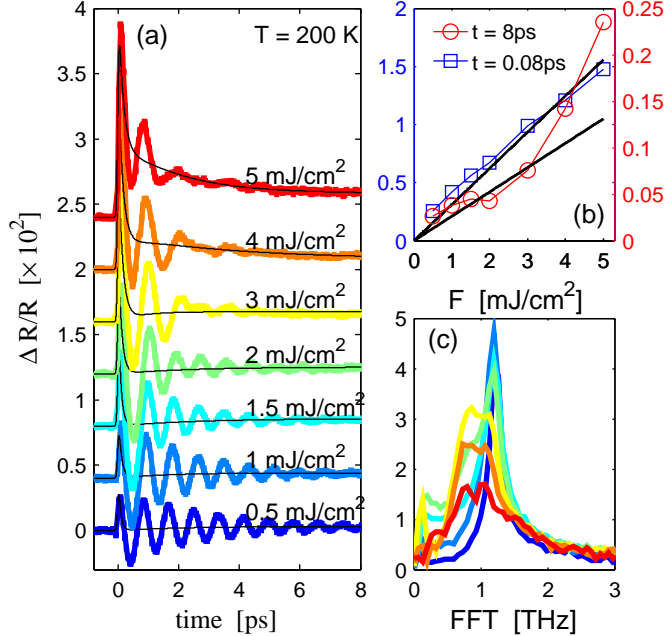}
\caption{(Color Online) a) Transient reflectivity curves for
different fluences at sample temperature T = 200~K. The solid lines
are a fit with $\Delta R_e$. b) $\Delta R/R$ at t = 0.08 and 4~ps.
Black lines are linear fits. c) Fast fourier transform of the
oscillatory part of the curves from a, applying the same color
scale. } \label{figure4}
\end{figure}


The phonon modes of the MT phase are also intimately related to the
structure, which as discussed is not unambiguously clarified in the
literature. X-ray diffraction data from the crystals used in this
study is consistent with an adaptive modulation. These results which
will be reported elsewhere confirm the 12 fold splitting of a
[202]$_c$ Bragg peak in the MT \cite{Kaufmann2010} and the existence
of all intermediate superlattice peaks corresponding to a
$(3\overline{2})_2$ structure \cite{Ustinov2009}. In a nanotwinned
structure the observed low frequency coherent phonons are truly
zone-folded acoustic phonons, similar to the coherent oscillations
observed in superlattices \cite{Bargheer2004}. At the Brillouin zone
center, the frequency 2d/v of the first mode is roughly given by the
speed of sound v divided by the period of the nanotwin super cell d.
The factor of 2 arises because a single nano-twinned cell has is
bound on both sides. Using elastic constants calculated for the
tetragonal structure of the MT phase \cite{Li2011}, we find
v$_{[110]}$ = 6.07~km/s for the mainly longitudinal wave, close to
the 5.54~km/s in the AUS phase \cite{Worgull1996}. The period of the
nanotwin super cell is d $\approx$ 21~$\AA$ \cite{Kaufmann2010}
giving a frequency of 1.4~THz. Considering the uncertainty of the
material constants and that we apply continuum theory in the
microscopic limit, this is in good agreement with the frequencies
found in \figref{figure2}. The phase of the oscillation is $\phi$ =
5.6 rad, which corresponds to a cosine oscillation starting $\sim$
100~fs after time zero. This suggest a displacive excitation, and
the slight delay with respect to time zero could indicate that the
driving force of the phonon mode is thermal strain
\cite{Thomsen1986} with the coherent phonon only being excited once
the electrons and lattice starts to thermalize. Since the structure
of Ni$_2$MnGa in the MT is debated an alternative explanation of the
observed coherent phonon is possible. If a periodic modulation of
the tetragonal unit cell does exist, the observed phonon could be an
amplitudon as for the PMT. The existence of a phason, which is the
acoustic variant of the optical amplitudon, has previously been
reported \cite{Shapiro2007}. With optical reflectivity we cannot
distinguish with certainty between the two explanations, but the
structure of the samples as determined by x-ray diffraction, and the
accuracy of the estimated frequency strongly supports the hypothesis
of a zone-folded acoustic phonon.

We finally consider the possibility of inducing the martensitic
phase transition with a fs laser pulse on an ultrafast timescale. In
\figref{figure4}a we show the transient reflectivity curves obtained
in the MT at T = 200~K just below T$_{MT}$ as a function of pump
fluences from 0.5 to 5 mJ/cm$^2$. The amplitude of the initial peak
at t = 0.08~ps is linear in fluence as shown in \figref{figure4}b,
where the solid lines are fits to the data. This linear dependence
is consistent with an electronic contribution from the laser heated
electron gas, with a temperature proportional to the fluence. In
contrast the change in reflectivity at t = 8~ps shows a highly
non-linear fluence dependence, with a clear increase in $\Delta$R
above $\sim3$~mJ/cm$^2$. This threshold behavior is one indication
of a laser induced phase transition, which for Ni$_2$MnGa at T =
200~K must be attributed to the martensitic transition. To confirm
this we consider in detail the transient reflectivity curves in
\figref{figure4}a. At low fluences ($<$ 2~mJ/cm$^2$) the amplitude
and the damping of the zone folded acoustic mode both increase with
fluence. At higher fluences a beating of the signal indicates the
presence of a second frequency. To extract the frequencies we
isolate the oscillatory part of the curves by subtracting the
incoherent part given by $\Delta R_e$. The remaining oscillating
signal is Fourier transformed as shown in \figref{figure4}c. At low
fluences the result is a single frequency at $\sim 1.2$~THz as
expected. As the fluence is increased a second peak gradually arises
at $\sim$ 0.8~THz. This frequency is consistent with the frequencies
found for the amplitudon in the PMT. It shows that with increasing
laser fluence the martensitic transition is induced, resulting in a
change of structure from MT to PMT, with the modulation of the PMT
realized by the amplitudon. The presence of a full period of the
zone-folded acoustic phonon at all applied fluences is a clear proof
that the nanotwinned structure remains present at early times (t $<$
2~ps). The beating on the other hand reveals that a significant
fraction of the PMT is present after 2~ps and confirms the expected
phase coexistence of PMT and MT. We have previously observed phase
coexistence in a laser induced first order phase transition
\cite{Mariager2012}.

To summarize we have measured two coherent optical phonons in the
pre-martensite and martensite phases of the Heusler alloy Ni$_2$MnGa
respectively. These were characterized as an amplitudon in the PMT
and a zone-folded acoustic mode arising from the nano-twinned
structure in the MT. The good estimate of the frequency of the
zone-folded acoustic mode supports the idea of an adaptive
martensitic phase in Ni$_2$MnGa. The martensitic phase transition
can be induced on a timescale of $\sim~2$~ps ps by laser excitation
and exhibits phase coexistence between the MT and PMT phases. A
confirmation of our findings will require further experiments. This
can be done by measuring the optical branches of the phonon
dispersion curves with inelastic neutron scattering, or by using
time-resolved x-ray diffraction to directly measure the atomic
motion in the time domain.

\begin{acknowledgments}
We thank M. Medarde for help with the magnetization measurements.
This work was supported by the Swiss National Foundation through
NCCR MUST.
\end{acknowledgments}


\end{document}